\documentclass[12pt,a4paper]{article}

\usepackage{amsmath}
\usepackage{amstext}
\usepackage[bookmarksopen=true]{hyperref}
\usepackage[a4paper,left=3cm,right=2cm,top=2.5cm,bottom=2.5cm]{geometry}
\usepackage[utf8]{inputenc}
\usepackage{graphicx}
\usepackage{multirow}
\usepackage{authblk}

\usepackage{listings}
\title{Simulating nonlocal no-signaling boxes}
\author[1]{Mátyás Koniorczyk}
\author[1]{Péter Naszvadi}
\author[1]{András Bodor}
\author[1,2]{Ottó Hanyecz}
\author[1,3]{Peter Adam}
\author[4]{Miklós Pintér}
\affil[1]{Department of Quantum Optics and Quantum Information, Institute for Solid State Physics and Optics, Wigner Research Centre for Physics, Budapest, Hungary}
\affil[2]{Department of Computer Algebra, Faculty of Informatics, Eötvös Loránd University, Budapest, Hungary}
\affil[3]{Institute of Physics, University of P\'ecs, P\'ecs, Hungary}
\affil[4]{Corvinus Center for Operational Research, Institute of Advanced Studies, Corvinus University of Budapest, Hungary}
\date{December 12, 2022.}

\begin{document}
\maketitle

\begin{abstract}
  We present a computer framework to simulate two-party boxes that exhibit nonclassical
  no-signaling correlations through a Web-based application
  programming interface (RESTful Web API). Unlike real quantum-based
  correlations, the simulated ones are not instantaneous and are
  created via communication with a trusted server. They can, however,
  be useful in a number of applications, including e.g. teaching the
  use of nonlocal correlations, designing and testing
  infocommunication systems, and engineering software interfaces for
  new quantum hardware. We demonstrate the use of the API via the
  simple implementation of the Clauser-Horne-Shimony-Holt game. Up to
  our knowledge no such a framework has been implemented or proposed
  thus far.
\end{abstract}

\section{Introduction}

The study of nonclassical correlations was triggered by the
Einstein-Podolsky-Rosen paradox~\cite{PhysRev.47.777} raising a
fundamental question of physics. The problem was first quantified by
J.S. Bell~\cite{PhysicsPhysiqueFizika.1.195} who studied a scenario of
two separated parties are in hold of a physical system each, so that
the two systems had interacted before. He pointed out that if the
parties can choose between different measurements on their systems,
the measurement results can show correlations that cannot be explained
by the assumption of pre-shared randomness; this is reflected in the
violation of certain inequalities. The underlying physical phenomenon
is quantum entanglement~\cite{RevModPhys.81.865, 2017}. Notably the
correlations obey the no-signaling property: they cannot be used for
transmitting information between the parties. The first experiment to
verify such correlations was proposed by Clauser, Horne, Shimony, and
Holt~\cite{PhysRevLett.23.880}, however, it was an extremely hard task
to produce such correlations with the technology of the 1960s.

The evolution of lasers and nonlinear optics in the 1990's, notably
the availability of entangled photon pairs~\cite{PhysRevLett.75.4337}
has brought Bell-type correlations to the forefront of research
interest. The structure of quantum and generic nonlocal no-signaling
correlations has been broadly studied and
understood~\cite{RevModPhys.86.419}. Device-independent quantum
cryptography, based on this kind of correlations, is now one of the
most promising technologies, and a broad variety of protocols have
been designed and demonstrated for numerous tasks, including secure
key distribution~\cite{2021QKD}, bit
commitment~\cite{2016dibitcommit}, or digital
signatures~\cite{2017sig}. Quantum communication with satellites
became now reality~\cite{PhysRevLett.115.040502}, and quantum
communication networks are being built~\cite{Simon_2017}.

In spite of this tremendous development, quantum technologies are
still expensive and not broadly accessible. While it is possible to
understand various protocols based on their descriptions, an actual
implementation leads to a deeper understanding. In addition, as the
technology is still in an experimental phase, the current physical
implementations are strongly tied to physical details, making them
less transparent from the viewpoints of a system engineer, a
programmer, or a mathematician. In fact, nonlocal no-signaling
correlations, as a mathematical notion, can also be understood without
any reference to any physical realization or motivation, following a
device independent approach entirely based on the mathematical
description of the behavior of ``black boxes''.  The possible
communication or cryptoprotocols based on such correlations are
interesting~\emph{per se}, and can be useful even when implemented
with simulated correlations.

In the present work we introduce a framework for simulating nonlocal
no-signaling correlations with certain limitations. As it is based on
the encrypted communication with a trusted server it cannot of course
operate between really (that is, space-like) separated parties and the
readout after the inputs is not instantaneous. On the other hand it
does not suffer from certain limitations that physical implementations
are facing, notably those of persistence. Moreover, it can simulate
not only quantum correlations, but also supraquantum ones: those
beyond quantum mechanical (i.e. physical) realizability.

As for the technical implementation, our service relies on RESTful WEB
API technology; the dominant one in network services currently. A
software library can be easily developed in virtually any development
environment or programming language that hides the otherwise simple
details of low-level API operation. This facilitates the
implementation, development, and testing of any protocol based on
nonlocal no-signaling correlations, the development of computer
applications using such resources, etc.

The aforementioned library can be easily modified to use a physical
device's API instead of the web-service based simulation. It is likely
that if the quantum technology to physically implement certain
nonlocal no-signaling correlations will mature, the physical devices
will practically appear in a way similar to our present implementation
to a software developer.  In this way an application developed using
our simulation framework can be easily modified to use physical
hardware in the future as quantum communication devices become
prevalent and affordable. Currently, on the other hand, it enables the
development and testing of protocols without the need of the currently
costly or not-yet-existent devices which can be readily converted to
use new physical hardware as soon as it becomes actually available.

This paper is organized as follows. Section~\ref{sec:nlboxes} gives a
brief introduction to the theory of no-signaling boxes. In
Section~\ref{sec:simul} we address the fundamental aspects of
simulating no-signaling boxes, including some details which become the
most apparent when a protocol implemented in
practice. Section~\ref{sec:implem} describes the system architecture
and the key implementational details. Section~\ref{sec:demo} provides
examples of the use of the service. In Section~\ref{sec:conclusions}
conclusions are drawn and an outlook is provided.

\section{Nonlocal boxes}
\label{sec:nlboxes}

Consider two parties, Alice and Bob, who are physically separated from
each other so the communication between them is excluded, apart
possibly from the following. They have access to a device (or, more
precisely, a pair of devices) which generates pairs of random variates
$(a,b)\in \mathcal{A}\times \mathcal{B}$ so that the variate $a$ is
available only for Alice while the other variate $b$ is available only
for Bob. Each output pair depends on an input pair, too, so that
Alice's input $x\in \mathcal{X}$ is entered by Alice locally, and so
is Bob's $y\in \mathcal{Y}$. The distribution of the output variates
depends on the pair of (local) inputs
$(x,y)\in \mathcal{X} \times \mathcal{Y}$ according to the conditional
probability distribution $P(a,b|x,y)$. Such devices will be termed as
a ``pairs of boxes'', or simply a ``box'' in what follows. We will
assume the sets $\mathcal{A}$, $\mathcal{B}$, $\mathcal{X}$,
$\mathcal{Y}$ to be finite. So far we allow for arbitrary
correlations; many bipartite boxes would enable communication between
the parties, though we will focus on those which do not in what
follows.

When a box is used multiple times\footnote{We note here that in some
  works a given transaction, i.e. a single use of a box is referred to
  as an instance of a box. When compared to those works, a ``box''
  there is a ``transaction'' in our terminology, whereas a ``box''
  there is a ``type of our box'' here.}, the input pairs and the
corresponding variate pairs have to be labeled. The labels $k$ will be
elements of an arbitrary index set $\mathcal K$, and the tuple
$(a_k,b_k,x_k,y_k)$ will be termed as the data of \emph{transaction}
$k$. We do not prescribe any ordering on the set $\mathcal{K}$, albeit
in practical realizations it is frequently related to time, e.g. due
to a causal ordering. We assume a Markovian property of the pair of
boxes, namely, the probability distribution of the variates
$(a_k,b_k)$ in a given transaction $k$ is entirely determined by
$x_k$, $y_k$, and $P(a,b|x,y)$.

Let us now restrict our attention to those boxes, which cannot be used
for the parties to communicate. This is the exclusion of signaling: it
implies that Alice and Bob cannot use their box to implement a
communication channel solely by using the boxes. In mathematical terms
this can be expressed with the following no-signaling conditions:
\begin{equation}
  \label{eq:ns1}
  \sum_b P(a,b|x,y) = P(a|x)\quad  \forall y,
\end{equation}
and similarly
\begin{equation}
\sum_a P(a,b|x,y) = P(b|y)\quad \forall x.
\end{equation}
Note that these conditions imply the existence of local marginals of
the joint conditional probability distribution. Hence, it is possible
to operate the boxes asynchronously: Alice can provide $x_k$ anytime,
obtaining $a_k$ immediately, and the same holds for Bob, $y_k$ and
$b_k$. The times of when a party use the box in a given transaction,
and thus the order of the uses is independent. This property can also
give rise to interesting protocols~\cite{PhysRevA.106.012223}. In what follows
we will restrict ourselves to no-signaling boxes.

\begin{figure}
  \centering
  \includegraphics[width=0.55\textwidth]{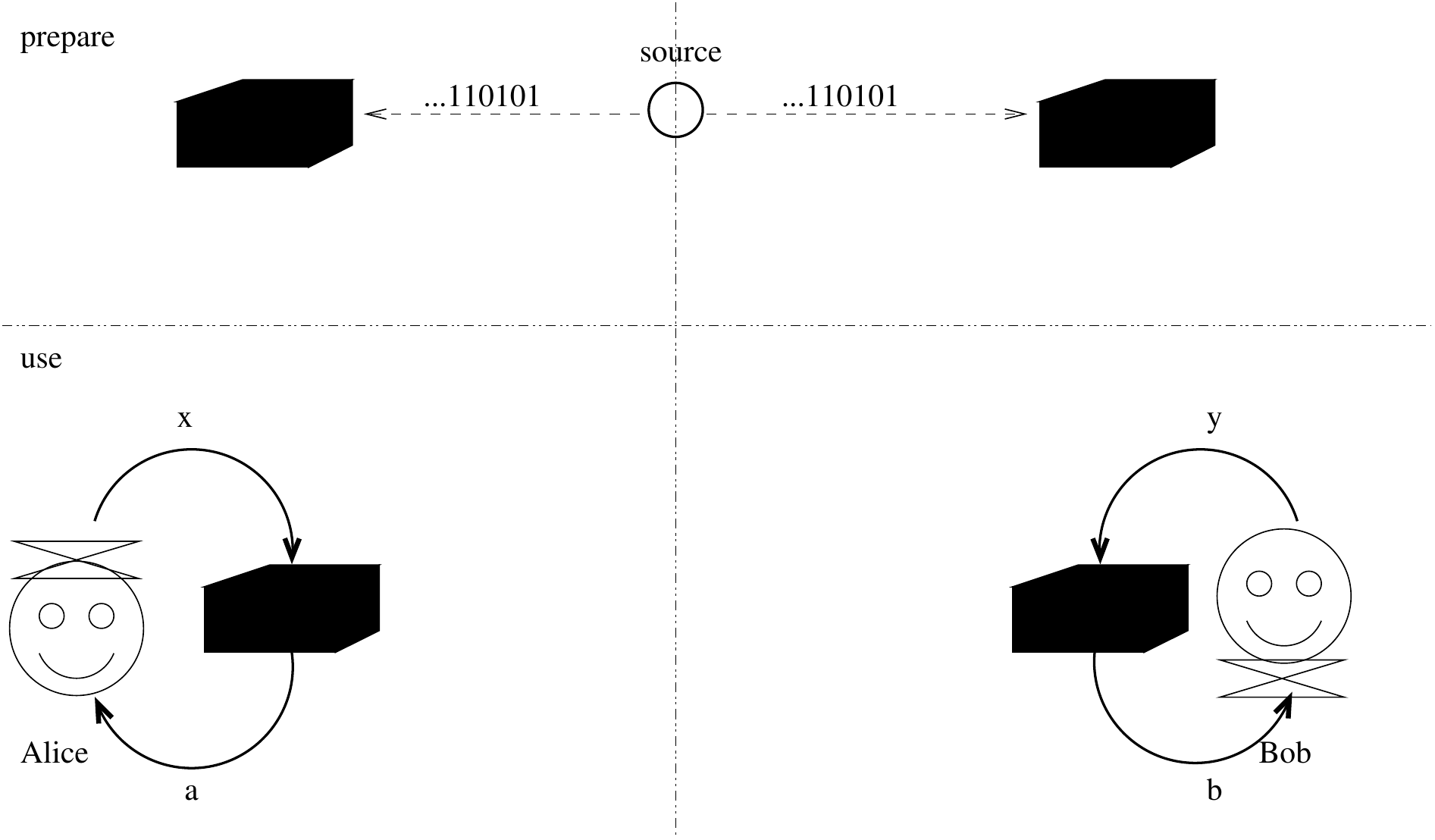}
  \caption{A local box pair: it can be implemented with randomness shared in
advance. The vertical dash-dotted line represents spatial separation,
whereas the horizontal one represent difference in time. Thus there
are two phases: the preparation of the box and the actual use. In the
second phase no communication is allowed between the boxes.}
\label{fig:clbox}
\end{figure}

The notion of locality of a box pair is to consider those
which can be realized with randomness shared in advance before the
transaction. A scenario with such a box pair is illustrated in Fig.~\ref{fig:clbox}.
Such boxes are described by conditional probability distributions that
can be expressed as a convex combination of products of local
deterministic boxes. A local deterministic box on Alice's side 
assigns a given $a(x)$ to each $x$, whereas such a box at Bob's side
assigns a given $b(y)$ to each $y$; their product is the parallel application
of the two. Such a pair of boxes has a deterministic (Dirac)
conditional probability distribution. Randomness shared in advance
enables the realization of any convex combination of these
distributions without any communication between the parties. Such
boxes are termed as ``local''.

No-signaling boxes form a significantly larger set than that
of the local boxes. Therefore there exist ``nonclassical
correlations'' which are interesting both fundamentally and in
applications, and some of these can be realized with physical
arrangements (i.e. quantum mechanically). Our simulation will address
nonlocal no-signaling boxes in general. Boxes that can be
realized physically include local boxes as a proper subset,
and they are a proper subset of no-signaling boxes. The structure
of the set of physically realizable boxes is defined by the laws of
quantum mechanics, we will not go into detail but will show an example
of this kind.

\begin{figure}
  \centering
  \includegraphics[width=0.55\textwidth]{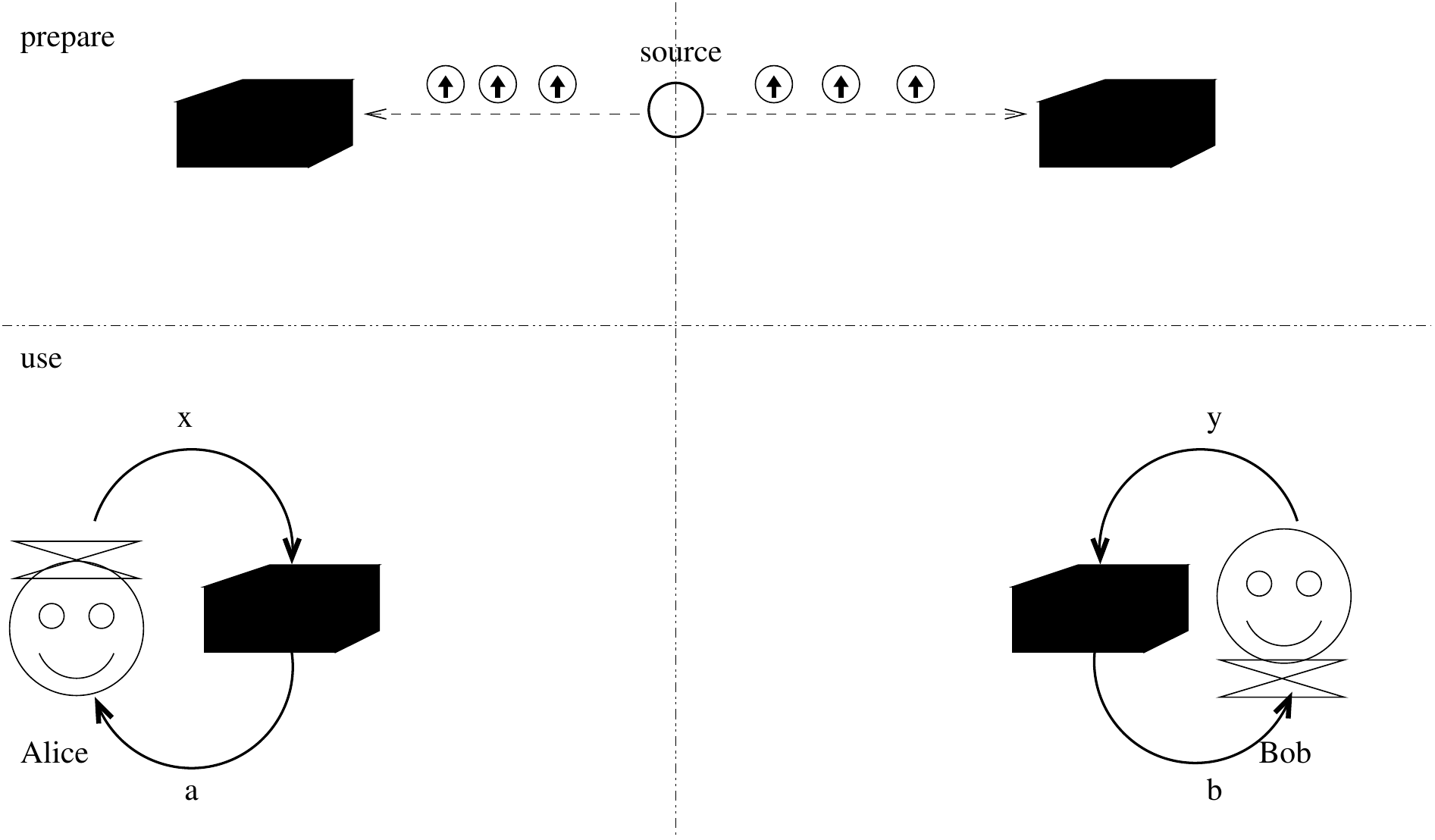}
  \caption{A quantum box pair, compare also with
    Fig.~\ref{fig:clbox}. The circles with upwards arrows inside
    represent quantum systems (e.g. particles); they are shared in
    advance. Due to their interaction at the source, they form pairs
    which are entangled. This enables the realization of nonlocal
    no-signaling correlation, albeit not the most general ones.}
  \label{fig:qbox}
\end{figure}

It can be expected that no-signaling boxes based on quantum
entanglement will be prevalently available as a
technology. Quantum-based boxes require the sharing of physical
systems, but there is no interaction between the elements of the box
during their operation. This scenario is depicted in Fig.~\ref{fig:qbox}.

The physical systems are particles; typically photons in case of many
realizations. The parts of the system are initially at the same
location, a source, and they are interacting, which results in an
entangled state in some of their internal degrees of freedom, like the
polarization of the two photons. The two subsystems are then sent to
the parties Alice and Bob, who choose the measurements corresponding
to the inputs $x$ and $y$ and carry them out on their particle to get
$a$ and $b$ as the measurement result.

Notably, the operation is instantaneous: the parties obtain the
(correlated) results immediately after sending the input to the box on
both sides, even if the parties' separation is spacelike, and thus
there is no way to communicate. Whenever Alice and Bob carry out their
measurements, the results are readily available right after the
completion of the measurement of each party; there is no need to
wait the minimum time that would allow the two sites to
communicate\footnote{Recall that information can only be propagated at
  a limited speed. There will be a ``local'' reaction time of the box,
  but this can be negligible.} This feature can be important in
certain applications~\cite{LaMura2005}.

The measurement by each party is done solely on the particle available
to the given party. Thus there is no interaction or communication
between the boxes at the parties (after sharing the pair of
particles). Nor there is any interaction or communication between the
boxes of Alice and Bob.  Thus it is guaranteed that no other party
will know about the particular values of $x(y)$ and $a(b)$ but Alice
(Bob).

Note also that from the no-signaling principle it follows that the two
parties may carry out their respective measurements anytime, in
arbitrary causal order, without synchronization.  If Alice and Bob
could store the particles for an arbitrarily long time, they could
share enough entangled particle pairs in advance and they could choose
freely when to make a given measurement. In practice, however, the
coherence times of such particles is short, thus the entangled state
is destroyed within a very short time. Hence, in practical scenarios
they obtain the particles from a central source (e.g. via fibers or
free-space propagation), and often there is a time synchronization to
ensure that the measurements are associated with actual members of
pairs. Hence, the realization of arbitrary timings, that is, using up pairs
with different timing and ordering deliberately on the two sides has up to our knowledge not yet been explored in experiments, although it would not be impossible, apart from some challenges due to loss and decoherence.

Nonlocal no-signaling boxes realized with quantum systems are
important for a number of proposed and already existing applications,
including protocols of quantum- and device independent
cryptography~\cite{2021QKD}, game theory~\cite{PhysRevA.101.062115}, etc.  Altogether, currently
the preparation of quantum-based box pairs is possible but is still an
experimental and technological challenge.  Hence, in order to collect
experience with the use of such boxes, it is reasonable to create a
simulator that can be used to experiment with them.

\section{Simulating no-signaling boxes}
\label{sec:simul}

Our goal is to simulate a pair of nonlocal no-signaling boxes whose
behavior is described by a given conditional probability distribution
$P(a,b|x,y)$. We define first what we mean by a logical implementation
or simulation as opposed to the physical realizations. Then we describe the
principle of the actual simulation algorithm.

\subsection{Logical implementation}

\begin{figure}
  \centering
  \includegraphics[width=0.55\textwidth]{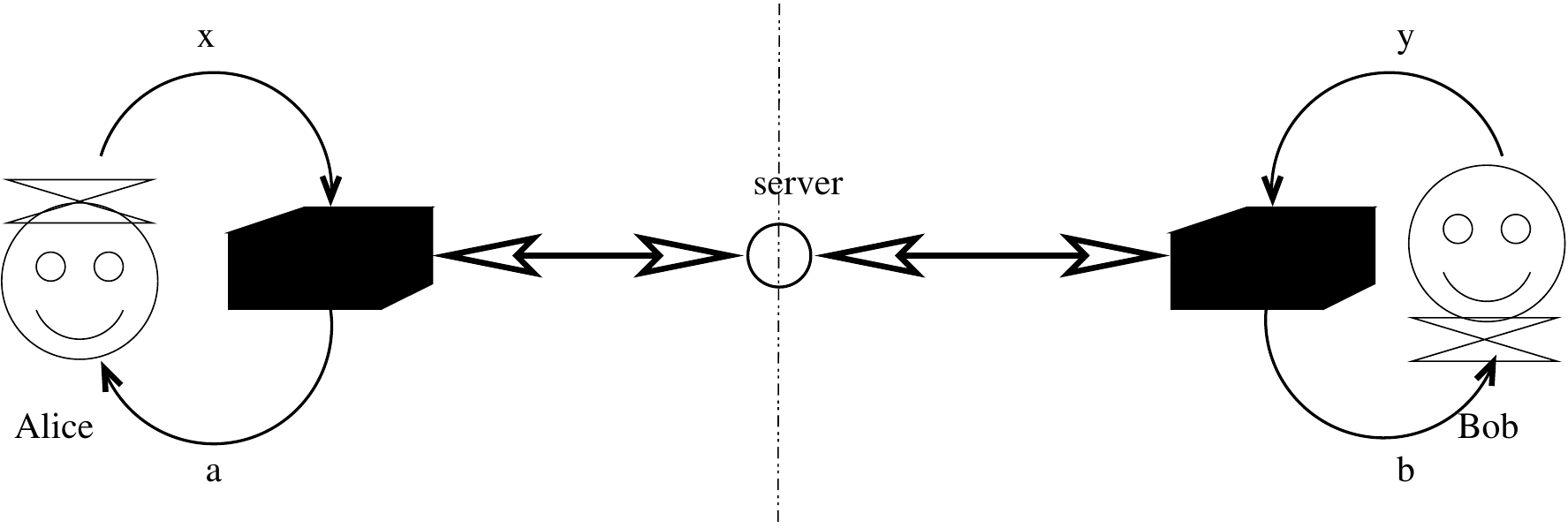}
  \caption{The nonlocal box emulation scenario. A bidirectional communication is allowed between \emph{the boxes} of Alice and Bob, via an encrypted channel, with a trusted server, during the whole process of using the box. Meanwhile Alice and Bob are still not able to communicate each other by using the box.}
  \label{fig:box_emu}
\end{figure}

In our simulation scenario we accept that there is two-way
communication between \emph{the boxes} at the parties and a trusted
server.  This is depicted in Fig~\ref{fig:box_emu}.
We require, however, that while the boxes themselves communicate, the
parties cannot use the box pair for sending any information: the
simulated correlations are no-signaling from the actors' perspective. Otherwise speaking, the \emph{correlations} themselves are nonlocal, regardless of the implementation.
Hence, our simulation scenario can well be considered also as a
\emph{logical implementation} of a nonlocal box.

When comparing with the physical implementation, as a trivial
consequence of being generated via communication the central server
will have all information about the results, and also there is a need to
wait for the completion of the communication with the central server
before the result becomes known, so the formation of the correlations
is certainly not instantaneous.  On the other hand, because of the
no-signaling principle, no synchronization assumed and the set of the
transactions $\mathcal{K}$ does not need to have a causal
structure. As we will point out later, this feature is easily
implemented in this framework.

As an additional benefit, it is certainly possible to simulate
correlations that are no-signaling but cannot be realized quantum
mechanically, such as a Popescu-Rohrlich (PR) box that will be described in Section~\ref{sec:demo} in detail. Assuming that the server is trusted,
the communication between the server and the boxes is secure, and that
the parties use the boxes according to the prescription, such an
implementation can be interesting \emph{per se}. We conclude this subsecetion with tabulating the required resources and the features of the various implementations in Tables~\ref{tab:comp_resource} and~\ref{tab:comp_feature}.

\begin{table}
  \centering
  \begin{tabular}{|l|c|c|c|}
    \hline
    type&local&quantum&logical\\
    \hline
    shared randomness&yes&no&no\\
    entanglement&no&yes&no\\
    bidirectional secure communication&no&no&yes\\
    \hline
  \end{tabular}
  \caption{A comparison of resources needed to realize a local, a
    quantum, and a logical (emulated) no-signaling box pair.}
  \label{tab:comp_resource}
\end{table}
\begin{table}
  \centering
  \begin{tabular}{|l|c|c|c|}
    \hline
    type&local&quantum&logical\\
    \hline
    instantaneous&yes&yes&no\\
    interaction-free&yes&yes&no\\
    quantum confidential&yes&yes&no\\
    quantum behaviors&no&yes&yes\\
    supraquantum behaviors&no&no&yes\\
    \hline
  \end{tabular}
  \caption{A comparison of features offered by a local, a
    quantum, and a logical (emulated) no-signaling box pair.}
  \label{tab:comp_feature}
\end{table}

\subsection{Algorithm}
\label{sec:boxemu_algorithm}

Let us now describe the actual algorithm of the simulation. Assume
first that Alice is the first to send her input, that is, she uses her
box with the given transaction in time before Bob. (Recall that no
synchronization is assumed but the transaction is uniquely identified
by a value of $k$.)  So Alice sends a particular $x_k$ value in
transaction $k$ to the box. The result $a_k$ of the box is drawn
according to the local marginal
\begin{equation}
  \label{eq:alicemarg}
P(a|x=x_k)=\sum_b P(a,b|x=x_k,y=\overline{y})
\end{equation}
where $\overline{y}\in \mathcal{Y}$ is an arbitrary fixed $y$ (due to
the no-signaling condition in Eq.~\eqref{eq:ns1} any element can be
chosen).  The respective value of the random variate $x_k$ is sent
to Alice, while the triple $(k,x,a)$ is stored in the database.

If Bob provides his input $(k,y_k)$ later and asks for his output
$b_k$, it is a random variate drawn according to the conditional
distribution
\begin{equation}
  \label{eq:bobmarg}
P(b|a=a_k,x=x_k,y=y_k)=\frac{P(b,a_k|x=x_k,y=y_k)}{P(a_k|x=x_k,y=y_k)},
\end{equation}
where
\begin{equation}
  \label{eq:bobmargnorm}
  P(a_k|x=x_k,y=y_k) = \sum_b P(b,a=a_k|x=x_k,y=y_k).
\end{equation}
and the transaction is completed (after storing all details in the
database).  As the protocol is symmetric, when Bob is the first to
initiate transaction $k$, the roles are reverted but the procedure is
the same.

In a software implementation of this simulation it is therefore vital
to ensure the following condition. When transaction $k$ has been
initiated by Alice, no reply to Bob can be generated before the
transaction has concluded for Alice, that is, before $a$ is generated
and $(k,x_k,a_k)$ has been stored. The same holds for Bob's initiation
of transaction $k$ for $(k,y_k,b_k)$. Using conventional relation
database management, this can be ensured by locking the table of
transactions, or at least transaction $k$ whenever it is acted upon on
behalf of either of the parties.

Note that there can be two kinds of actions: if the transaction was
already initiated by the other party then we use the joint probability
with the known condition, whereas if it wasn't we just use the local
marginal but keep the given input. Looking at the empirical marginals
ex post, they will follow the local marginal distributions that exist
because of the no-signaling condition.

\section{Implementation}
\label{sec:implem}

\begin{figure}
  \centering
  \includegraphics[width=0.6\textwidth]{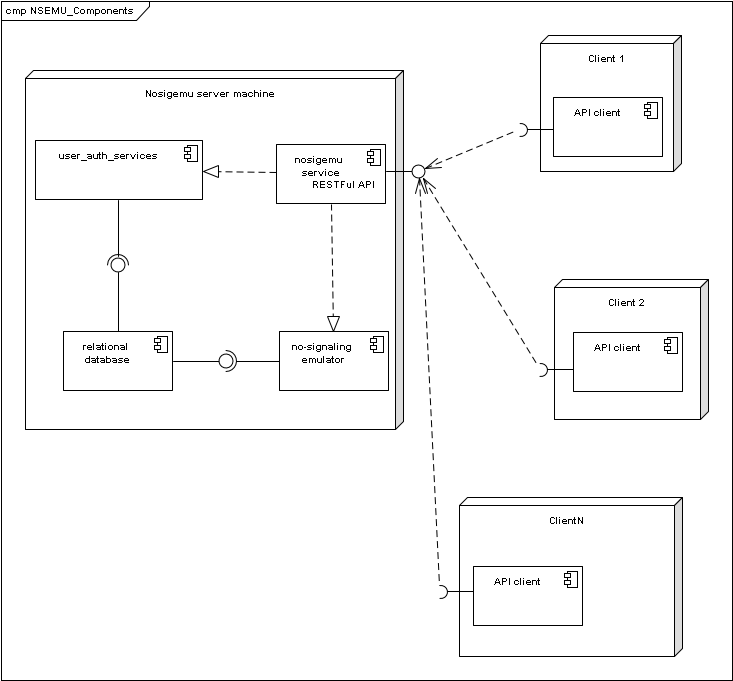}
  \caption{UML\cite{UML} component diagram of the software architecture.}
  \label{fig:component}
\end{figure}

In this Section we describe the software architecture that has been
used for the implementation. The components of the IT architecture are
depicted in Figure~\ref{fig:component}. The simulation is based on a
central service run on a server. The service provides a RESTful API to
clients, using HTTP GET requests with URL parameters, and returning
the result in JSON format. (An example of a session will be presented
later.)

The server component realizes a component needed for user
authentication and management, and a component that realizes the PR
box emulator described in Section~\ref{sec:boxemu_algorithm}.  Both of
the components use the same underlying relational database which they
communicate via its standard internal interface.

The server component is implemented in Python programming language.
It is based on SQLAlchemy~\cite{sqlalchemy} as an object-relational mapper and Flask~\cite{grinberg2018flask} as the WEB API provider
framework. The currently running beta version uses PostgreSQL~\cite{postgresql} as a
relational database manager. The random variates used by the server at
the time of writing this paper are obtained from a "Quantis" USB
Quantum Random Number Generator, model ``USB-4M'', manufactured by "ID
Quantique"\cite{QRNG} with the serial number 184443A410. The Python
library for accessing this device was also developed in the framework
of the present project\cite{QRNGlib}. At the time of the publication
of this article as an e-print, the beta version will be available for the
public after a free registration, for academic and educational purposes~\cite{NLBOXweb}.

\begin{figure}
    \centering
    \includegraphics{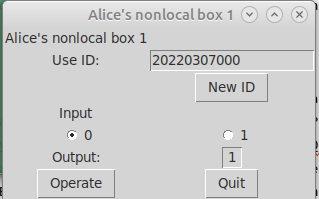}
    \caption{A simple desktop graphical user interface for the logical nonlocal box implementation}
    \label{fig:qrnd}
\end{figure}

Owing to the use of a standard API, a client can be any device
running any software that is capable of consuming RESTful APIs at a
basic level. 
Hence the possible client implementations and devices
range from tutorial codes in various programming languages through
smartphone applications to test crypto protocol
implementations. A screenshot of a simple desktop graphical user interface is to be found in Fig.~\ref{fig:qrnd}

\section{Demonstration}
\label{sec:demo}

In this section we demonstrate the use of simulated no-signaling
correlations by implementing the so-called Clauser-Horne-Shimony-Holt (CHSH) game~\cite{PhysRevLett.23.880, PhysRevA.101.062115}. It can be
considered as a tutorial project with pairs of participants who
implement their box pair, and then play the game with and without
using it.

The game itself is the following.  The two players, Alice and Bob are
separated and are not allowed to communicate. In each turn of the game
Alice randomly chooses an input $x$ which is 0 or 1, while Bob
randomly chooses an input $y$ which is 0 or 1. Importantly, they have
to be really honest about choosing these with a uniform
distribution. Alternatively they can be provided these inputs by a
trusted source. Then Alice says an output $a$, Bob says an output $b$.
They both get a unit of reward in the following two cases: if both of
them chose $1$ as input ($x=y=1$), and their output is the opposite,
i.e. $a=1$, $b=0$ or $a=0$, $b=1$, or if any of them had 0 as an
input, and their outputs $a$ and $b$ are same. Otherwise there is a
unit of negative payoff. The payoff function is thus the same for both
parties; it is tabulated in Tab.~\ref{tab:chsh}.
\begin{table}
  \centering
  $$
  \begin{array}{|lll||cc|cc|}
    \hline
    x\downarrow& y\rightarrow&&\multicolumn{2}{|c}{0}&  \multicolumn{2}{c|}{1}\\
    &a\downarrow& b\rightarrow& 0  & 1 &  0  & 1  \\
    \hline 
    \hline
    \multirow{2}{*}{0}&0  & &  1 & -1  & 1  & -1    \\
    \cline{2-3}&1&&            -1  & 1 & -1  & 1   \\
    \hline
    \multirow{2}{*}{1}&0  & &  1  & -1  & -1 & 1   \\
    \cline{2-3}&1&&              -1  & 1  & 1  & -1  \\
    \hline
  \end{array}
  $$
  $$
  \begin{array}{|lll||cc|cc|}
    \hline
    x\downarrow& y\rightarrow&&\multicolumn{2}{|c}{0}&  \multicolumn{2}{c|}{1}\\
    &a\downarrow& b\rightarrow& 0  & 1 &  0  & 1  \\
    \hline 
    \hline
    \multirow{2}{*}{0}&0  & &  1/2 & 0 & 1/2  & 0    \\
    \cline{2-3}&1&&            0 & 1/2 & 0  & 1/2   \\
    \hline
    \multirow{2}{*}{1}&0  & &  1/2 & 0 & 0 & 1/2    \\
    \cline{2-3}&1&&            0  & 1/2  & 1/2  & 0  \\
    \hline
  \end{array}
  $$
\caption{The payoff function of the CHSH game (top) and the PR-box, the
no-signaling behavior maximizing it (bottom).}
\label{tab:chsh}
\end{table}

This game is the so-called Clauser-Horne-Shimony-Holt (CHSH) game,
which is behind the celebrated CHSH inequality. It can be proven that
if Alice and Bob are not allowed to communicate and are restricted to
use pre-shared randomness (even an infinite sequence of correlated
random bits, shared before their separation), the best they can do is
the following: they agree in advance to always say $a=0$ ($b=4$)
regardless of their inputs $x$ and $y$. In that case they will win in
75\% of the cases, that is, in case of uniformly distributed input
pairs they get 1 in 3/4 of the cases and -1 in 1/4 of them, so the
average payoff will be 1/2. It can be proven that no other, even
randomized strategy involving pre-shared randomness can result in a
better payoff. This limit on the payoff is the Bell-CHSH inequality.

The limit of $1/2$ on the average payoff can be overcome when the
parties are allowed to use a pair of no-signaling boxes. The nonlocal
no-signaling behavior of a box pair that enable Alice and Bob to
obtain the maximal average payoff of 1, is also tabulated in
Tab.~\ref{tab:chsh}; is called the Popescu-Rohrlich box. If they both
feed their box with with their inputs $x$ and $y$ and they provide the
respective output $a$ and $b$ as the output, they will be positively
rewarded in all the cases. It can be shown, however, that the access to
such a box pair does not able to send any message or signal to the
other. They get , however, coordinated without communication.

Let us now see how the gameplay is actually implemented using the API
calls.  (We will use the ~curl~ command available on most Linux systems to
communicate the API with GET request. Alternatively, the URL can be
written into the browser.)
\begin{enumerate}
    \item Alice sends $x=0$ as her input to box 1, a CHSH box. The transaction id is a date followed by a 3-digit zero-padded ordinal number.
    \begin{lstlisting}
    curl --get 'https://nonlocalbox.wigner.hu/api/v1/useBox?boxID=1&transactionID=20211106001&x=0&apiKey=$ALICE_KEY' 
    
    {"a":1,"boxID":1,"status":0}
    \end{lstlisting}
    The box has emitted the reply $a=1$. The zero {\tt status} implies that there is no error.
    \item Let now Bob send $y=0$. Note that for $x=y=0$ the results should be correlated, so Bob should get $b=1$. And indeed, 
    \begin{lstlisting}
    curl --get 'https://nonlocalbox.wigner.hu/api/v1/useBox?boxID=1&transactionID=20211106001&y=0&apiKey=$BOB_KEY'
    {"b":1,"boxID":1,"status":0}
    \end{lstlisting}
    \item In a next transaction (with an incremented ID), Bob will be the first to send $y=1$:
    \begin{lstlisting}
    curl --get 'https://nonlocalbox.wigner.hu/api/v1/useBox?boxID=1&transactionID=20211106002&y=1&apiKey=$BOB_KEY'

    {"b":1,"boxID":1,"status":0}
    \end{lstlisting}  
    The box gave $b=1$. 
    \item Now assume that Alice also opts for $x=1$, thus the results should be anticorrelated, i.e. $a=0$ should be obtained. And indeed:
    \begin{lstlisting}
    curl --get 'https://nonlocalbox.wigner.hu/api/v1/useBox?boxID=1&transactionID=20211106002&x=1&apiKey=$ALICE_KEY'

    {"a":0,"boxID":1,"status":0}
    \end{lstlisting}
\end{enumerate}
Note that both parties obtain a uniformly distributed random result
for their inputs, when observed just locally. However, when analyzed
together, the expected joint conditional probability of the
CHSH nonlocal box can be observed.  We encourage the reader to
register for the service and verify this.

\section{Conclusions}
\label{sec:conclusions}

We have reported on the design and implementation of a RESTful WEB API service that implements nonlocal no-signaling correlations logically. Thereby it is capable of simulating nonlocal quantum correlations that are perhaps the  most intriguing features of quantum mechanics and are essential ingredients of most applications in quantum information and communication, notably in device independent quantum cryptography. The described web service has also been implemented by us and we made it available to the community~\cite{NLBOXweb}.
 
 From the point of view of scientific research, one of our contributions is the algorithm that implements no-signaling correlations using a central trusted resource: we have not seen it before in the literature. The discussion of the asynchronous nature of no-signaling correlations can also be considered as a minor contribution of this kind. While it is mentioned in some previous contributions, probably because of its difficult implementation in quantum experiments, it gained less attention before. During the development we report here, this has lead us to finding the first known application in which the non-sequential use of nonlocal no-signaling resources is useful~\cite{PhysRevA.106.012223}.
 
 From the technological point of view our contribution makes nonlocal no-signaling correlations readily available using the perhaps most commonly used web service technology. This can serve as a test and development environment for applications, even those designed for physical realizations of quantum correlations; device independent cryptographic protocols for instance. The API technology paves the way of designing a broad range of applications ranging from demonstrations on various platforms as well as practically useful ones, possibly.
 
 Finally, from the dissemination point of view we believe that our API is an enabler in the experience-based teaching of nonclassical correlations and Bell-type quantum phenomena. At the time of writing of this paper, this application of our API is being tested in a high-school environment, and a mobile phone application is planned to increase the dissemination impact. We hope that these will help a number of people to understand the basics of the phenomena whose experimental study has lead to the Nobel prize in Physics awarded in 2022~\cite{Nobel}. 
\section*{Acknowledgments} 
This research was supported by the National Research, Development, and
Innovation Office of Hungary under project numbers K133882 and K124351
and the Ministry of Innovation and Technology and the National
Research, Development and Innovation Office within the Quantum
Information National Laboratory of Hungary and the Thematic Excellence
Programme. M.K. wants to thank L\'aszl\'o Pere for inspiring
discussions.


\end{document}